\documentclass[twocolumn]{revtex4}

\usepackage{graphicx}
\usepackage{amsmath}
\usepackage{bbold}
\usepackage{xcolor}
\usepackage{dsfont}
\usepackage{mathrsfs}
\newcommand\myfrac[2]{\genfrac{}{}{0pt}{}{#1}{#2}}

\begin{document}

\title{Anisotropic deformation of the 6-state clock model: Tricritical-point classification}  

\author{Maria Polackova$^{1}$ and Andrej Gendiar$^2$}\email[Corresponding author:\ ]{andrej.gendiar@savba.sk}

\affiliation{$^{1}$Department of Physics, Imperial College London, London, SW7 2AZ, United Kingdom}

\affiliation{$^2$Institute of Physics, Slovak Academy of Sciences, D\'{u}bravsk\'{a} cesta 9, 845 11 Bratislava, Slovakia}

\begin{abstract}
The two-dimensional $q$-state clock models exhibit the Berezinskii-Kosterlitz-Thouless (BKT) transition for $q\geq5$ since they are a subset of the isotropic XY model. We examine the $6$-state clock model with an anisotropic deformation. Selecting the $6$-state Potts model as a source of the deformation, the model naturally violates the discrete rotational symmetry of the clock model. We introduce the anisotropic deformation parameter $\alpha$ in the clock model interpolating the clock ($\alpha = 1$) and the Potts ($\alpha = 0$) models. We employ the corner transfer matrix renormalization group method to analyze the phase transitions on the square lattice in the thermodynamic limit. Three different phases and phase transitions are identified. The phase diagram is constructed, and we determine a tricritical point at $\alpha_{\rm c} = 0.21405(4)$ and $T_{\rm c} = 0.834017(5)$. Analyzing the latent heat and the entanglement entropy in the vicinity of the $T_{\rm c}(\alpha_{\rm c})$, we observe a single discontinuous phase transition and two BKT phase transitions meeting in the tricritical point. The tricritical point exhibits a phase transition of the second order with the critical exponents $\beta \approx 1/10$ and $\delta \approx 14$. We conjecture that an infinitesimal surrounding of the tricritical point consists of the three fundamental phase transitions, in which the first and the BKT orders gradually weaken into the second-order tricritical point.
\end{abstract}



\maketitle

\section{Introduction}

The multi-state spin models provide sufficient information about various types of phase transitions. For classical statistical systems, phase transitions of the first, second, and infinite order reflect the microscopic properties of spin models. The higher number of spin states and richer spin interactions are essential in our study of phase transitions.

Theoretical models simplify realistic systems and usually do not satisfactorily describe real physical properties observed experimentally. A solution could be to determine missing sources responsible for discrepancies between theory and experimental measurements. This work proposes a novel approach. We use multi-state spin models to enhance irregular behavior, rather than to include additional terms into Hamiltonians to search for new sources of physical effects. We propose a way of modifying (deforming) spin interactions in simplified spin models to introduce an anisotropic deformation breaking a spin-rotational symmetry. Among the typical spin models, we select the two-dimensional $q$-state clock model in which we define the anisotropic spin deformation with respect to the $q$-state Potts spin interaction.

This work was inspired by experimental observations of the spin anisotropies where an external magnetic field induced magnetic anisotropy in FeSiB thin films~\cite{an1}. Also, experimentally confirmed organometallic compounds yield a spin anisotropy affecting a rotational symmetry of the two-dimensional spin system where, for weak spin anisotropies, there are Heisenberg-Ising and Heisenberg-XY crossovers for the finite-temperature phase diagrams, respectively, with two Ising and a single topological Berezinskii–Kosterlitz–Thouless (BKT) phase transitions~\cite{an2}. Anisotropies violate the regular behavior of the simplified models, as experimentally evident in~\cite{an3, an4, an5}.

The $q$-state clock models exhibit the second-order (continuous) phase transition if $q=2,3,4$, however, for $q\geq5$ there are two infinite-order Berezinskii-Kosterlitz-Thouless (BKT) phase transitions in the $q$-state clock model~\cite{XY1, XY2, XY3, XY4, BKT1, BKT2, BKT3, clock1}. The $q$-state Potts models show second-order (continuous) phase transitions if $q=2,3,4$ whereas first-order (discontinuous) phase transitions occur if $q\geq5$~\cite{FYWu, Potts5, PhysA, Po1, Po2, Po3}. If sorted into the universality classes, the $2$-state Ising universality and the $3$-state Potts universality are common for both models. However, the $4$-state Potts universality with logarithmic corrections differs from the Ising universality of the $4$-state clock model which decouples into the two Ising models yielding again the Ising universality. The correlation length diverges ($\xi \to \infty$) for the second-order transitions, whereas the first-order transitions result in the finite $\xi < \infty$.

There is a dramatic change in the phase-transition nature between $q=4$ and $q=5$. For instance, long-range correlations of the $5$-state Potts model weaken the typical first-order phase transitions due to the proximity of the second-order transition ($q=4$). Hence, the discontinuous transition at $q=5$ is similar to the continuous transition. As a consequence, the correlation length, $\xi_q$, becomes extraordinary large ($\xi_5 \approx 2512$) at the phase transition, although it is  finite~\cite{Potts5}. If compared to the Potts models with $q \geq 6$, the correlation length shortens, i.e., $\xi_6 \approx 159$, $\xi_7 \approx 48$, etc.~\cite{Sandwick}. Therefore, numerical analysis is more sensitive to the discontinuous transitions for larger $q$, where less accuracy is sufficient at the phase transition. On the other hand, the clock model with $q\geq5$ is known for the existence of an intermediate critical phase, which separates the low-temperature ferromagnetic phase from the high-temperature paramagnetic one. The intermediate phase is the BKT phase. Again, the proximity of the second-order phase transition at $q=4$ makes the $5$-state clock model difficult to treat more accurately since the BKT phase is rather narrow, as shown later.

This study brings a different insight into the form of spin interactions. We consider anisotropic deformations of the discrete rotational symmetry in the clock models. The deformed spin interactions thus lead to a novel phase structure of spin systems. Motivated by the phase transitions' complexity, we propose an anisotropic multi-state spin model, where first-order and two infinite-order (BKT) transitions are allowed to interfere, resulting in a specific tricritical point. It exhibits different nature as the tricritical point typically emerged in the Blume-Capel (BC) model~\cite{BCm1, BCm2, BCm3, BCm4, BCm5, BCm6} where first- and second-order transitions meet.

We employ a numerical method, Corner Transfer Matrix Renormalization Group (CTMRG) \cite{Nishino, CTMRG, APS, TMRG, DMRG, CTMRG1, CTMRG2}, which is an accurate numerical tool capable of determining the tricritical point. The appropriateness of the CTMRG method is supported by our earlier work~\cite{TCp} in which we analyzed the $J_1$-$J_2$ spin model by CTMRG, where the second-order transition gradually weakens and transforms into a discontinuous first-order transition at a tricritical point, analogously to the BC model.

The anisotropic interactions included in the $q$-state clock model violate its discrete rotational symmetry by the Potts-like interactions. The two Potts and clock models combined together in this work give rise to a so-called PC model. It mixes the first- and infinite-order transitions resulting in the tricritical point. The analysis of how the two distinct BKT transitions merge and gradually change into the second-order transition in the tricritical point goes beyond the scope of this study and will be studied elsewhere. The accurate location of the BKT phase transition requires an additional scaling, even for the simple 6-state clock model~\cite{BKT1}. In order to understand this infinite-order transition in deeper detail, the anisotropic deformation is helpful because it gradually weakens the BKT transitions. The simplest case is a linear interpolation between the infinite-order and the first-order phase transitions.

We primarily focus on the latent-heat calculation below the tricritical point where the first-order transition is present. On the other hand, we observe the formation of the BKT phase by calculating the entanglement entropy and the specific heat. The size of the BKT phase shrinks if approaching the tricritical point from the opposite direction. Our intention is to locate and classify the phase-transition order at the tricritical point.It is worth mentioning that the two BKT infinite-order transitions resulting in the clock models with q>4 require extensive computational resources to be determined adequately by CTMRG~\cite{BKT1}. On the other hand, the first-order and the second-order transitions are easier to be treated numerically by CTMRG. Hence, we examine the tricritical point in the PC model with satisfactory numerical accuracy.

The paper is organized as follows.
In Sec.~2, we define the $q$-state spin models and construct the phase diagram of the Potts and clock models. A concise introduction to the CTMRG method is given in Sec.~3, where we summarize thermodynamic quantities. Section~4 is devoted to numerical results, where we first calculate the entanglement entropy and latent heat when studying the discontinuous and the BKT transitions, respectively. We construct a $T(\alpha)$ phase diagram of the model and locate the tricritical point $T_{\rm c}(\alpha_{\rm c})$. Finally, we classify the tricritical point by calculating its critical exponents $\beta$ and $\delta$. The anisotropic deformation of the model is concluded in Sec.~5.

\section{Spin model}  

Consider $q$-state spins $\sigma_{i,j}=0,1,\dots, q-1$ and each spin $\sigma_{i,j}$ lies on the two-dimensional square lattice of the infinite size at position $i$ and $j$. The nearest-neighbor spin pairs interact via $J$. The spins can enter the $q$-state Potts-model Hamiltonian
\begin{equation}
	{\mathscr{H}}_{\rm total}^{\,q-{\rm P}} =-J\hspace{-0.2cm}\sum\limits_{i,j=-\infty}^{+\infty}
 	 \delta_{(\sigma_{i,j};\sigma_{i+1,j})}
	+\delta_{(\sigma_{i,j};\sigma_{i,j+1})}
\end{equation}
and the $q$-state clock-model Hamiltonian
\begin{equation}
    {\mathscr{H}}_{\rm total}^{\,q-{\rm C}} =-J\hspace{-0.2cm}\sum\limits_{i,j=-\infty}^{+\infty}\hspace{-0.2cm}
    \cos(\theta_{i,j} - \theta_{i+1,j})  
    +\cos(\theta_{i,j} - \theta_{i,j+1}) \, ,
    \label{clock}
\end{equation}
where $\theta_{i,j} = \frac{2\pi}{q}\sigma_{i,j}$.
The entire square lattice is built up of horizontally and vertically interacting spin pairs. When $q=2$, both models belong to the Ising universality class (including the $4$-state clock model)~\cite{Baxter}. Within the models, the phase transitions of the first, second, and infinite (BKT) orders can be accessed by varying $q$.

\begin{figure}[tb]
\includegraphics[width=0.48\textwidth]{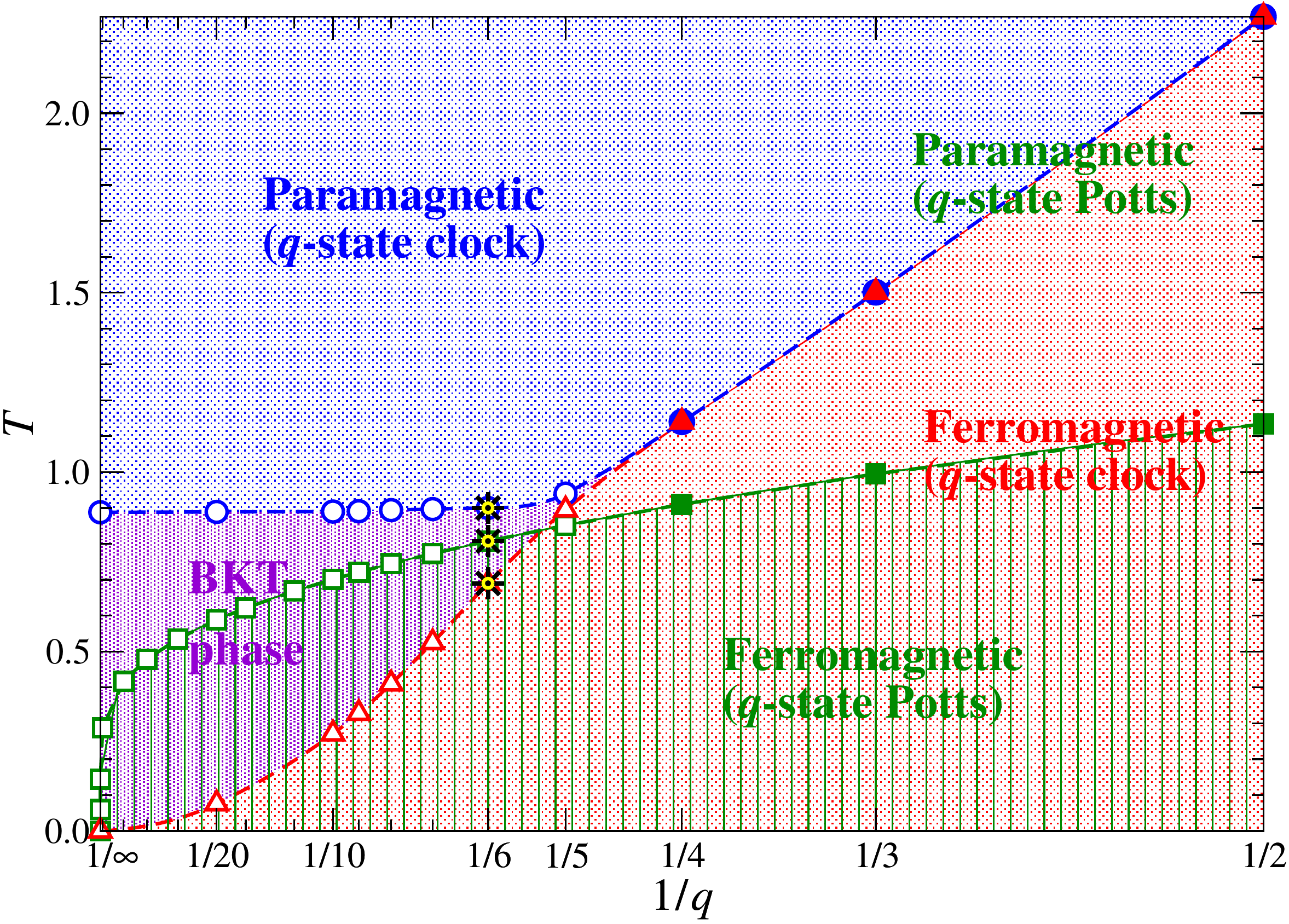}
\caption{The phase diagrams (temperature $T$ vs inverse $q$) for both the Potts and the clock models joined into one graph. The squares denote the discrete phase-transition temperatures $T_0$ for the $q\geq5$-state Potts model. The circles (in blue) and triangles (in red), respectively, denote the transition temperatures $T_1$ and $T_2$ for the $q$-state clock model. The lines connecting the symbols serve as a guide for the eye. The filled symbols correspond to the second-order phase transitions $T_0$ and $T_1 \equiv T_2$ for both models when $q=2,3,4$ only. The three asterisks point out the current study carried out at $q=6$.}
\label{Fig1}
\end{figure}

Figure~\ref{Fig1} sketches the phase diagram of the $q$-state Potts and the $q$-state clock models in the ferromagnetic regime $J>0$. The phase diagrams of both models are plotted in the same graph using the CTMRG method for $2\leq q\leq20$. For $q > 20$, we used the analytic formula for the Potts model as the phase-transition temperature $T_0=1/\ln(1+\sqrt{q})$ are known exactly~\cite{FYWu}. The squares in the phase diagram separate the ferromagnetic phase (marked by the vertical green lines) from the paramagnetic phase above the phase-transition line. The $q$-state clock model exhibits a BKT phase located between the low-temperature ferromagnetic phase and the high-temperature paramagnetic phase. This is true for finite $q\geq5$. Taking the limit $q\to\infty$ leads to the suppression of the ferromagnetic phase in both cases. The clock model with $q\to\infty$ describes the classical XY model in Eq.~\eqref{clock} with continuous spin $0\leq \theta < 2 \pi$~\cite{XY1,XY2,XY3,XY4}.

Let the total Hamiltonian depend on a number of the spin states $q$ and be parameterized by a real number $\alpha$ such that $0\leq\alpha\leq1$. The parameter $\alpha$ linearly interpolates between the Potts model if $\alpha=0$ and the clock model if $\alpha=1$. The thermodynamic properties are governed by the partition function
\begin{align}
	\nonumber
	{\cal Z}_{q}(\alpha)
	= & \sum\limits_{\myfrac{\rm spin}{\rm configs.}}
	\exp{\left(-\frac{{\mathscr{H}}_{\rm total}^{\,q{\rm-PC}}
	(\alpha)} {k_{\rm B}T}\right)} \\
	= & \sum\limits_{\myfrac{\rm spin}{\rm configs.}}
	\prod\limits_{i,j=-\infty}^{\infty}
	\exp\left(-\frac{H^{q{\rm-PC}}_{i,j,\vartheta} (\alpha)} {k_{\rm B}T}\right)\,,
    \label{Z_qa}
\end{align}
where the summation runs over all spin configurations of the total Hamiltonian $\mathscr{H}_{\rm total}^{\,q{\rm-PC}} (\alpha)$ on the infinite square lattice. Boltzmann constant and thermodynamic temperature are assigned to $k_{\rm B}$ and $T$, respectively. Since we consider only the nearest-neighbor interactions between the classical spins, the total Hamiltonian can be factorized into the sum of local Hamiltonians $H^{q{\rm-PC}}_{i,j,\vartheta} (\alpha)$. We choose the local Hamiltonian with three spins. Let the indices $i$ and $j$, respectively, denote the X and Y axis coordinates on the square lattice. The three spins $\sigma_{i,j}$, $\sigma_{i,j+1}$, and $\sigma_{i+1,j}$ form the local Hamiltonian and contribute with one vertical spin pair and one horizontal spin pair sharing the spin $\sigma_{i,j}$ on the lattice site $i$ and $j$.

\begin{figure}[tb]
\includegraphics[width=0.48\textwidth]{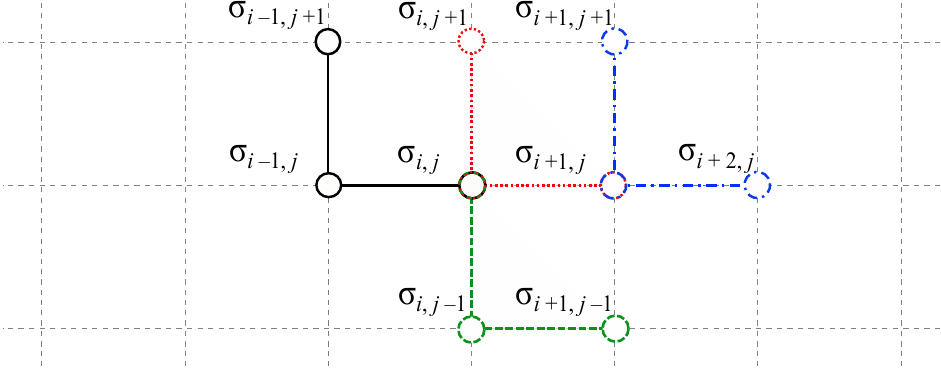}
\caption{Example of positioning $4$ three-spin local Hamiltonians $H^{q{\rm-PC}}_{i-1,j,\vartheta} (\alpha)$, $H^{q{\rm-PC}}_{i,j,\vartheta} (\alpha)$, $H^{q{\rm-PC}}_{i+1,j,\vartheta} (\alpha)$, and $H^{q{\rm-PC}}_{i,j-1,\vartheta} (\alpha)$ on the square lattice (thin dashed lines). The circles denote the spins $\sigma$ and the horizontal and vertical lines correspond to the nearest-neighbor interactions $J$.}
\label{Fig2}
\end{figure}

We consider ferromagnetic coupling ($J>0$) and impose a uniform homogeneous magnetic field $h$ aligned with a chosen spin direction $\vartheta$, typically, $\vartheta = 0$. Employing the linear interpolation means that we can express the three-spin local Hamiltonian in the following
\begin{align}
	H^{q{\rm-PC}}_{i,j,\vartheta}(\alpha) = 
        \sum\limits_{k=0}^{1}
	(1-\alpha){\cal H}_{i,j,k,\vartheta}^{q-{\rm Potts}}
	+ \alpha\,{\cal H}_{i,j,k,\vartheta}^{q-{\rm clock}}\, .
	\label{local_H_alpha}
\end{align}
The last line expresses the local Hamiltonian by the Potts part ${\cal H}_{i,j,k,\vartheta}^{q-{\rm Potts}}$ and the clock part ${\cal H}_{i,j,k,\vartheta}^{q-{\rm clock}}$ separately. Three-spin local Hamiltonians $H^{q{\rm-PC} }_{i, j, \vartheta} (\alpha)$ covers the entire square lattice, see Fig.~\ref{Fig2} (if the lattice is finite, periodic boundary conditions are useful).

In addition, we introduced index $k$ to split the three-spin local Hamiltonian between one spin pair oriented vertically ($k=0$) and another oriented horizontally ($k=1$). We have chosen this form to keep the vertical and horizontal Hamiltonians symmetric and identical, i.e.,
${\cal H}_{i,j,0,\vartheta}^{q-{\rm Potts}} \equiv {\cal H}_{i,j,1,\vartheta}^{q-{\rm Potts}}$.
For the Potts model, we get
\begin{align}
	\nonumber
	{\cal H}_{i,j,k,\vartheta}^{q-{\rm Potts}}
	= - & J\, \delta_{(\sigma_{i,j};\,\sigma_{i+k,j+1-k})} \\
	  - & \frac{h}{4}
	  \left[
	  \delta_{(\sigma_{i,j},\vartheta)} + \delta_{(\sigma_{i+k,j+1-k},\vartheta)}
	  \right]\,.
\end{align}
The clock model also satisfies ${\cal H}_{i, j, 0, \vartheta}^{q-{\rm clock}} \equiv
{\cal H}_{i,j,1, \vartheta}^{q-{\rm clock}}$
and
\begin{align}
	\nonumber
	{\cal H}_{i,j,k,\vartheta}^{q-{\rm clock}}
	= & -J\cos\left[\frac{2\pi}{q}\left(\sigma_{i,j}
	-\sigma_{i+k,j+1-k}\right)\right]\\
	 & - \frac{h}{4} \cos\left[\frac{2\pi}{q}
	\left(\sigma_{i,j}-\vartheta\right)\right] \\
	\nonumber
	 & - \frac{h}{4} \cos\left[\frac{2\pi}{q}
	\left(\sigma_{i+k,j+1-k,j}-\vartheta\right)\right]
	\,.
\end{align}

Selecting $q=6$, the $6$-state Potts model can be expressed in form of the $6\times6$ diagonal matrix
\begin{equation}
    \renewcommand{\arraystretch}{1.2}
    \renewcommand{\tabcolsep}{0.2cm}
	{\cal H}^{6-{\rm Potts}}_{i,j,k,0}
	= -\left(
	\begin{tabular}{rrrrrr}
		$\tilde{J}$ & 0 & 0 & 0 & 0 & 0 \\
		0 & $J$ & 0 & 0 & 0 & 0 \\
		0 & 0 & $J$ & 0 & 0 & 0 \\
		0 & 0 & 0 & $J$ & 0 & 0 \\
		0 & 0 & 0 & 0 & $J$ & 0 \\
		0 & 0 & 0 & 0 & 0 & $J$ \\
	\end{tabular}
	\right)
	\label{al0}
\end{equation}
for any $i,j,k$. The magnetic field $h$, acting in the direction $\vartheta=0$, is included in $\tilde{J}=J+\frac{h}{4}$. Analogously, the $6$-state clock model has the form

\renewcommand{\arraystretch}{1.2}
\renewcommand{\tabcolsep}{0.05cm}
\begin{equation}
	{\cal H}^{6-{\rm clock}}_{i,j,k,0}
	= -\frac{1}{2}\left(
	\begin{tabular}{rrrrrr}
		$2\tilde{J}$ & $\tilde{J}$ & $-\tilde{J}$ & $-2\tilde{J}$ & $-\tilde{J}$ & $\tilde{J}$ \\
		$\tilde{J}$ & $2J$ & $J$ & $-J$ & $-2J$ & $-J$ \\
		$-\tilde{J}$ & $J$ & $2J$ & $J$ & $-J$ & $-2J$ \\
		$-2\tilde{J}$ & $-J$ & $J$ & $2J$ & $J$ & $-J$ \\
		$-\tilde{J}$ & $-2J$ & $-J$ & $J$ & $2J$ & $J$ \\
		$\tilde{J}$ & $-J$ & $-2J$ & $-J$ & $J$ & $2J$ \\
	\end{tabular}
	\right).
	\label{al1}
\end{equation}
The full PC model parameterized by $\alpha$ in
Eq.~\eqref{local_H_alpha} gives 
\renewcommand{\arraystretch}{1.2}
\renewcommand{\tabcolsep}{0.055cm}
\begin{align}
    \nonumber
	{\cal H}^{6{\rm-PC}}_{i,j,k,0}(\alpha)
	= -\frac{J}{2}\left(
	\begin{tabular}{rrrrrr}
		$2$ & $\alpha$ & $-\alpha$ & $-2\alpha$ & $-\alpha$ & $\alpha$ \\
		$\alpha$ & $2$ & $\alpha$ & $-\alpha$ & $-2\alpha$ & $-\alpha$ \\
		$-\alpha$ & $\alpha$ & $2$ & $\alpha$ & $-\alpha$ & $-2\alpha$ \\
		$-2\alpha$ & $-\alpha$ & $\alpha$ & $2$ & $\alpha$ & $-\alpha$ \\
		$-\alpha$ & $-2\alpha$ & $-\alpha$ & $\alpha$ & $2$ & $\alpha$ \\
		$\alpha$ & $-\alpha$ & $-2\alpha$ & $-\alpha$ & $\alpha$ & $2$ \\
	\end{tabular}
	\right)
\end{align}
\renewcommand{\tabcolsep}{0.12cm}
\vspace{-0.3cm}
\begin{align}
	\hspace{1.95cm} -\frac{h}{8} \left(
	\begin{tabular}{rrrrrr}
		 $2$ & $\alpha$ & $-\alpha$ & $-2\alpha$ & $-\alpha$ & $\alpha$ \\
		$\alpha$ & $0$ & $0$ & $0$ & $0$ & $0$ \\
		$-\alpha$ & $0$ & $0$ & $0$ & $0$ & $0$ \\
		$-2\alpha$ & $0$ & $0$ & $0$ & $0$ & $0$ \\
		$-\alpha$ & $0$ & $0$ & $0$ & $0$ & $0$ \\
		$\alpha$ & $0$ & $0$ & $0$ & $0$ & $0$ \\
	\end{tabular}
	\hspace*{-0.12cm}\right).
	\label{Halpha}
\renewcommand{\tabcolsep}{0.0cm}
\renewcommand{\arraystretch}{1.0}
\end{align}

We can generalize the local Hamiltonian for any $q$-state PC model by $q \times q$ matrix. If we index the $q$-state matrix subscripts with $\sigma, \sigma^\prime = 0,1,2,\dots,q-1$, each matrix element has the expression
\begin{align}
	\nonumber
	\left[{\cal H}^{q{\rm-PC}}_{i, j, k, \vartheta}(\alpha)\right]_{\sigma,\sigma'}
	 = - \ & \alpha J\cos\left[\frac{2\pi}{q}(\sigma-\sigma')\right] \\
	 \nonumber
	 - \, & \frac{\alpha h}{4} \cos\left[\frac{2\pi}{q}(\sigma-\vartheta)\right]
	 \delta_{(\sigma';\vartheta)} \\
	 \nonumber
	 - \, & \frac{\alpha h}{4} \cos\left[\frac{2\pi}{q}(\sigma'-\vartheta)\right]
	 \delta_{(\sigma;\vartheta)} \\
	 \nonumber
	 +\, & ( \alpha - 1)\left(J + \frac{h}{2}\delta_{(\sigma;\vartheta)}\right)
	 \delta_{(\sigma;\sigma')} \\
	+ &  \frac{h}{4}\delta_{(\sigma;\vartheta)}
	 \delta_{(\sigma;\sigma')} \, .
\end{align}
For brevity, we have omitted the lattice positions $i,j$ for both spins $\sigma$ such that we simplify them into $\sigma\equiv\sigma_{i,j}$ and $\sigma'\equiv\sigma_{i+k,j+1-k}$ for either $k$.

While varying $0 \leq \alpha \leq 1$ with a given $q \geq 5$, the first-order phase transition, characteristic for the Potts model ($\alpha=0$), gradually changes and, at some point, splits into two infinite-order (BKT) phase transitions, which are typical for the clock model ($\alpha=1$).

If there exists a point $\alpha_{\rm c}$, possibly a tricritical point, a continuous (second-order) phase transition is very likely at $\alpha_{\rm c}$. It is our task to locate the point $\alpha_{\rm c}$ and determine its critical exponents. There could be another scenario, according to which there is no continuous transition at $\alpha_{\rm c}$, and some discontinuities in thermodynamic functions may possibly occur.

In the following, we claim that with increasing $0 \leq \alpha \leq \alpha_{\rm c}$, the first-order transition gradually weakens when approaching the tricritical point $\alpha_{\rm c}$. At $\alpha_{\rm c}$, the first-order discontinuity vanishes and changes into the second-order phase transition. Right above $\alpha_{\rm c}$ the two infinite-order (BKT) transitions coexist and start moving away with increasing $\alpha > \alpha_{\rm c}$.

\section{CTMRG method}

The CTMRG method, developed by Nishino~\cite{Nishino}, is an alternative approach to the transfer-matrix method~\cite{TMRG}, which is meant to be the classical counterpart to the density matrix renormalization group method~\cite{DMRG} for one-dimensional quantum systems. CTMRG is designed to calculate the thermodynamic properties of the two-dimensional classical spin systems on the square lattice. It is an iterative algorithm based on subsequent density-matrix renormalizations. The square-shaped lattice is successively built-up by iterative expansions of four corner transfer matrices. The corner transfer matrix can be thought of as a generalized form of the transfer-matrix approach~\cite{Baxter}.

If integer $k=1,2,3,\dots,\infty$ enumerates the CTMRG iteration steps, then each of the four corner transfer matrices ${\cal C}_k$ successively expands its size $(k+1)\times(k+1)$. Hence, the four corner transfer matrices ${\cal C}_k$ (sharing the central spin site) form the entire square lattice with the total number of spins $N_k = (2k+1)^2$. At each iteration step $k$, the following recursion scheme is used to expand ${\cal C}_k$~\cite{Nishino}
\begin{align}
    {\tilde{\cal C}}_{k+1} & = {\cal W}_{\rm B}{\cal T}^2_k{\cal C}_k\, ,\\
    {\tilde{\cal T}}_{k+1} & = {\cal W}_{\rm B}{\cal T}_k\, ,
\end{align}
where ${\cal W}_{\rm B}$ is Boltzmann weight and ${\cal T}_k$ is a transfer matrix between two adjacent half-length rows/columns of spins. The iterative procedure is initialized with ${\cal C}_1 \equiv {\cal T}_1 \equiv {\cal W}_{\rm B}$. The Boltzmann weight is defined on the square-shaped four-spin plaquette $\{\sigma_{i,j}\sigma_{i+1,j}\sigma_{i,j+1}\sigma_{i+1,j+1}\}$. Having rescaled $J\to J/2$ and $h\to h/2$~\cite{APS} and $\tilde\beta=-1/(k_{\rm B}T)$, we get
\begin{equation}
    {\cal W}_{\rm B}=\exp{\left(\tilde\beta\sum\limits_{k=0}^{1}{\cal H}^{q{\rm-PC}}_{i, j, k, \vartheta}(\alpha)+{\cal H}^{q{\rm-PC}}_{i+k, j+1-k, 1-k, \vartheta}(\alpha)\right)}\,.
\end{equation}

The partition function ${\cal Z}_q (\alpha)$, from Eq.~\eqref{Z_qa}, can be directly calculated using the corner-transfer-matrix formalism~\cite{Nishino, Baxter}. We demonstrate it on the relation between the free energy ${\cal F}$ normalized per spin and the corner transfer matrix ${\cal C}_k$. In particular,
\begin{eqnarray}
\nonumber
{\cal F} (q,\alpha) & \hspace*{-0.2cm} = & \hspace*{-0.2cm} \lim\limits_{k\to\infty} {\tilde\beta}^{-1} N_k
\ln {\cal Z}_q (\alpha)\\
& \hspace*{-0.2cm} = & \hspace*{-0.2cm} \lim\limits_{k\to\infty} {\tilde\beta}^{-1} (2k+1)^2 \ln\left( {\rm Tr}\, {\cal C}_k^{\,4} (q,\alpha)\right) ,
\label{FreeEng}
\end{eqnarray}
where ${\rm Tr}\,{\cal C}_k^{\,4} (q,\alpha)$ is a tensor product of four combined corner transfer matrices forming the square lattice. Each corner transfer matrix ${\cal C}_k (q,\alpha)$ is, in fact, a rank-3 tensor $[{\cal C}_k(q,\alpha)]^{\sigma}_{\xi\zeta}$ with three indices; one of them is a $q$-state spin variable $\sigma$ and the two indices $\xi,\zeta$ describe multi-state ($m$-state) renormalized spin variables. Then, the partition function ${\cal Z}={\rm Tr} \left({\cal C}_k^{\,4} \right)$ has an expression (omitting $q$ and $\alpha$ for brevity)
\begin{equation}
    {\rm Tr}\left({\cal C}_k^{\,4} \right) = \sum\limits_{\sigma=1}^{q}
    \sum\limits_{\myfrac{\xi\ \xi^{\prime}}{\xi^{\prime\prime}\xi^{\prime\prime\prime}}=1}^{m}
    \left[{\cal C}_k\right]^{\sigma}_{\xi\xi^\prime}
    \left[{\cal C}_k\right]^{\sigma}_{\xi^{\prime}\xi^{\prime\prime}}
    \left[{\cal C}_k\right]^{\sigma}_{\xi^{\prime\prime}\xi^{\prime\prime\prime}}
    \left[{\cal C}_k\right]^{\sigma}_{\xi^{\prime\prime\prime}\xi}.
\end{equation}

The structure of CTMRG enables us to construct a reduced density matrix $\rho_k$ (being $qm\, \times\, qm$ matrix). At arbitrary iteration step $k$, we evaluate $\rho_k$ by partially tracing out tensor indices of the four corner transfer matrices~\cite{Nishino}
\begin{equation}
    {\left[\rho_k\right]}^{\sigma\xi}_{\tilde{\sigma}\tilde{\xi}} = 
    \sum\limits_{\xi{^\prime}\xi^{\prime\prime}\xi^{\prime\prime\prime}=1}^{m}
    \left[{\cal C}_k\right]^{\sigma}_{\xi\xi^{\prime}}
    \left[{\cal C}_k\right]^{\sigma}_{\xi^{\prime}\xi^{\prime\prime}}
    \left[{\cal C}_k\right]^{\tilde{\sigma}}_{\xi^{\prime\prime}\xi^{\prime\prime\prime}}
    \left[{\cal C}_k\right]^{\tilde{\sigma}}_{\xi^{\prime\prime\prime}\tilde{\xi}}.
\end{equation}

When applying CTMRG to studying spin systems, we need to find an expression for the reduced density matrix $\rho \equiv \rho_{k\to\infty}$ in the thermodynamic limit. The limit $k\to \infty$ has to be understood in terms of a sufficiently large iteration number $k_{\rm max}\gtrsim10^4$ above which all thermodynamic functions, normalized per spin, are fully converged. The convergence is always reached, regardless of whether the spontaneous symmetry breaking has occurred in the system or not. 

The numerical accuracy of CTMRG is controlled by the maximal spin configuration space. It is given by the cutoff dimension $m$ of the projection operator ${\cal P}_k$ (it is a unitary operator ${\cal P}^{\dagger}_k{\cal P}^{~}_k = \mathds{1}$). The dimension denotes the number of the spin states kept, and we assign it to the integer variable $m$~\cite{Nishino, DMRG}. The reduced density matrix has the dimension $qm\times qm$. The projection operator ${\cal P}_k$ consists of such eigenstates of the reduced density matrix which correspond to its $m$ largest eigenvalues $1\geq\omega_1\geq\omega_2\geq\cdots\geq\omega_m\geq\omega_{m+1}\geq\cdots\omega_{qm}\geq0$, i.e., the most probable spin configurations. Hence,
\begin{equation}
    \sum\limits_{r=1}^{qm}
    \sum\limits_{s=1}^{qm}
    \left[{\cal P}^{\dagger}_k\right]_{i,r} \left[\rho_k\right]_{r,s} \left[{\cal P}^{~}_k\right]_{s,j} = \omega_i \delta_{i,j}\, ,
\end{equation}
where $i,j=1,2,\dots,m$. The full eigenvalue spectrum is normalized, $\sum_{i=1}^{qm} \omega_i=1$, because we require normalization of ${\cal C}_k$ so that ${\rm Tr}\rho_k = 1$.

The projection operators ${\cal P}_k$ are expressed as rectangular matrices $qm\,\times\,m$ applied at each iteration $k$ to keep the exponentially expanding Hilbert space $(q^{N_k})$ computationally feasible. The restricted and renormalized dimension of the Hilbert space is constant $qm^4 \ll q^{N_k}$. Then, the tensors ${\tilde{\cal C}}_k$ and ${\tilde{\cal T}}_k$ of dimensions $q^3m^2$ and $q^4m^2$, respectively, shrink down to $qm^2$ and $q^2m^2$ while undergoing the renormalization transformations
\begin{align}
    {\cal P}^{\dagger}_{k} {\tilde{\cal C}}^{~}_{k} {\cal P}^{~}_{k} & \to {\cal C}^{~}_{k}\,.\\
    {\cal P}^{\dagger}_{k} {\tilde{\cal T}}^{~}_{k} {\cal P}^{~}_{k} & \to {\cal T}^{~}_{k}\,.
\end{align}

The higher the $m$, the higher the numerical accuracy of the thermodynamic functions monotonously converging to the exact solution is reached. To some extent, the numerical accuracy can be specified as a so-called truncation error
\begin{align}
\epsilon_m = \sum\limits_{i=m+1}^{qm} \omega_i\, ,
\end{align}
which should be $0\leq \epsilon_m \ll 1$ (we summed up the least-probable states related to the truncated eigenvalues $\omega_i$). The lowest numerical accuracy of CTMRG happens at phase transitions with the corresponding truncation error $\epsilon_m \approx 10^{-9}$ which is accessible for $m>150$ for this study. Not too far away from the phase transition, we get $\epsilon_m \approx 10^{-16}$ which is almost exact up to $15$ (machine precision) leading digits of the thermodynamic functions.

Ideally, all thermodynamic quantities are equivalent to exact solutions only if keeping $m=q^k$, which is not computationally feasible. However, when evaluating the thermodynamic limit ($N_k \to \infty$), we also need to extrapolate the thermodynamic quantities with respect to $m \to \infty$ in order to obtain accurate results.

We can evaluate the order parameter (spin magnetization $M$ of the Potts-model spins) in arbitrary spin direction $\vartheta$
\begin{equation}
M = \langle\sigma\rangle
= {\rm Tr}\left[\rho\,\delta_{(\sigma;\vartheta)}\right]\, ,
\label{Mg}
\end{equation}
The internal energy $U = {\cal F} - T\partial_T {\cal F}$, which is also equivalent to the nearest-neighbor spin-spin correlation,
\begin{equation}
U = -2J\langle\sigma\sigma^\prime\rangle = -2J\,{\rm Tr}\left\{\rho\cos\left[\frac{2\pi}{q}
(\sigma-\sigma^\prime)\right]\right\}\, .
\label{IntEng}
\end{equation}
The latent heat $\lambda$ is directly obtained from the internal energy by taking left and right limits at the phase-transition temperature $T_0$ 
\begin{equation}
\lambda = U_+ - U_- = \lim\limits_{T\to T_0^+}U(T) - \lim\limits_{T\to T_0^-}U(T)\, .
\label{LH}
\end{equation}
If the first-order phase transition is present, then $\lambda>0$. The specific heat (capacity)
\begin{equation}
c = \frac{\partial\,U}{\partial\, T} = -T\frac{\partial^2 {\cal F}}{\partial\, T^2}
\end{equation}
exhibits two peaks in the proximity of both BKT transitions. Finally, we calculate the von-Neumann entanglement entropy
\begin{equation}
S = - {\rm Tr}\left(\rho\ln\rho\right)=-\sum\limits_{j=1}^m \omega_j\ln\omega_j
\label{EE}
\end{equation}
by the eigenvalues $\omega_j$ of the reduced density matrix $\rho$.

The phase transitions are characterized by the singular behavior of the above thermodynamic quantities. Typically, they (or their derivatives) exhibit singular behavior with a sharp maximum. The maximum usually diverges in the thermodynamic limit $N\to\infty$ such as $S$ and $c$. Alternatively, we can observe discontinuities in phase transitions of the first order, in which those quantities are accompanied by either non-diverging maxima ($S$ and $c$) or non-analytic behavior, such as sudden jumps in $M$ and $U$.

We can uniquely identify the first-order phase transitions by calculating the nonzero latent heat $\lambda>0$ in Eq.~\eqref{LH}. Let us recall that the precise location of the discontinuous jump in the internal energy $U$ can be determined only by minimizing the free energy from a crossover caused by imposing open and fixed boundary conditions~\cite{APS}. Then, an abrupt discontinuity at phase transition $T_0$ also appears in the magnetization $M$ (measured in any direction $\vartheta=0,1, \dots q-1$), including the entanglement entropy $S$ and the specific heat $c$.

\section{Results}

We study the PC model within the entire interval $0\leq\alpha\leq1$ by thermodynamic quantities in Eqs.~\eqref{FreeEng} and \eqref{Mg}-\eqref{EE}. The first-order transition at temperature $T_0$ is characterized by a discontinuity in $M$, $U$, $S$, and $c$. The latent heat $\lambda > 0$, cf Eq.~\eqref{LH}, where ${\cal F}$ is non-analytic (exhibits a kink). Hence, left and right derivatives of ${\cal F}$ have to  differ at $T_0(0\leq \alpha < \alpha_{\rm c})$, i.e,
\begin{equation}
-T {\left.\frac{\partial{\cal F}}{\partial T}\right\vert}_{T \to T_0^+} = U_{+}
\ \ >
\ \ -T {\left.\frac{\partial{\cal F}}{\partial T}\right\vert}_{T \to T_0^-} = U_{-}.    
\end{equation}

On the other hand, the two infinite-order transitions at temperatures $T_1$ and $T_2$ can be identified by the typical shape in $S$ and two shifted divergent peaks in $c$. The spontaneous magnetization logarithmically converges to zero between $T_1$ and $T_2$ because the entire temperature region $T_1 \leq T \leq T_2$ is critical. The BKT transition observed for finite $m$ at $T_1$ is known to be underestimated whereas it is overestimated at $T_2$ for $\alpha=1$ only. Hence, additional scaling analyses are required by CTMRG~\cite{BKT1} and MC simulations~\cite{BKT3}.

From now on, we consider the case of $q=6$ only. The anisotropic deformation in the 6-state PC model starts with the detection of discontinuities in $M$, $S$, and $U$. Particularly, we investigate the limit $\lambda \to 0$, as $\alpha \to \alpha_{\rm c}$, where the discontinuous transition gradually turns into the continuous at $\alpha_{\rm c}$. On the other hand, we expect that the two BKT transitions at $T_1$ and $T_2$ start approaching one another if $\alpha$ decreases from $1$ to $\alpha_c$. The detailed classification of the phase transition at $\alpha_c$ is of our interest. We also localize the corresponding critical temperature $T_c$ at $\alpha_c$, for details, see Tab.~I. 

    \begin{table}[tb]    
    \renewcommand{\arraystretch}{1.3}
    \caption{Summarized characteristics of the transitions.}
    \begin{tabular}{lccc}
        \hline
        \hline
        Interval & $0\leq\alpha<\alpha_{\rm c}$ & $\alpha_{\rm c}$ & $\alpha_{\rm c}<\alpha\leq1$ \\
        \hline
        Notation & $T_0(\alpha)$ & $T_c(\alpha_c)$ & $T_1(\alpha)$, \, $T_2(\alpha)$ \\
        \hline
        Region& left & tricritical & right \\
        \hline
        Order & first & second & $2\times$ infinite \\
        \hline
        Transition & discontinuous & continuous & continuous \\
        \hline
        Latent heat & $\lambda>0$ & $\lambda=0$ & $\lambda=0$ \\
        \hline
        Trunc. error & $\epsilon_{200}\lesssim10^{-10}$ & $\epsilon_{200}\lesssim10^{-9}$ & $\epsilon_{200}\lesssim10^{-9}$ \\
        \hline
        \hline
    \end{tabular}
    \end{table}
    \renewcommand{\arraystretch}{1.0}

We, therefore, start with the construction of the $T(\alpha)$ phase diagram, in which we approximately determine the tricritical point $T_c(\alpha_c)$. The free-energy crossover specifies the first-order transition temperature $T_0$ in the entire interval $0\leq \alpha < \alpha_{\rm c}$. We evaluate the internal energy $U$ with the discontinuity at $T_0(\alpha)$ which is associated with the latent heat $\lambda$. Taking the limit $\lambda\to 0$ at $\alpha_c$ and $T_c$, we estimate the tricritical point $T_c(\alpha_c)$.

Further analysis improves the tricritical point $T_c(\alpha_c)$ including the exponents $\beta$ and $\delta$. The numerical accuracy expressed by the truncation error reaches the value of $\epsilon_{m=200} \lesssim 10^{-9}$. This value guarantees satisfactorily reliable results at the tricritical point $T_c(\alpha_c)$. In this work, we consider the number of states kept $m$ within the interval $50 \leq m \leq 200$.

We use dimensionless units and set $k_{\rm B}=1$. To simplify the problem, we consider the ferromagnetic ordering at low  temperatures and we set the interaction $J = 1$. As mentioned earlier, we introduce the parameter $0\leq\alpha\leq1$ to explore the gradual transformation between a discontinuous phase and the BKT phase transition. The PC model linearly interpolates between the Potts ($\alpha=0$) and the clock ($\alpha=1$) models. We can observe such nontrivial features only when $q\geq5$, see Fig.~\ref{Fig1}. The higher the $q$, the more computational cost is required to reach the desired accuracy. On the other hand, when $q=5$, the correct determination of the two BKT phase transitions encounters extensive numerical effort with high numerical accuracy to detect the two closely placed BKT phase transitions. We select $q=6$ in the following, as emphasized by the three asterisks in Fig.~\ref{Fig1}.
 
\subsection{Phase diagram}

\begin{figure}[tb]
\includegraphics[width=0.48\textwidth]{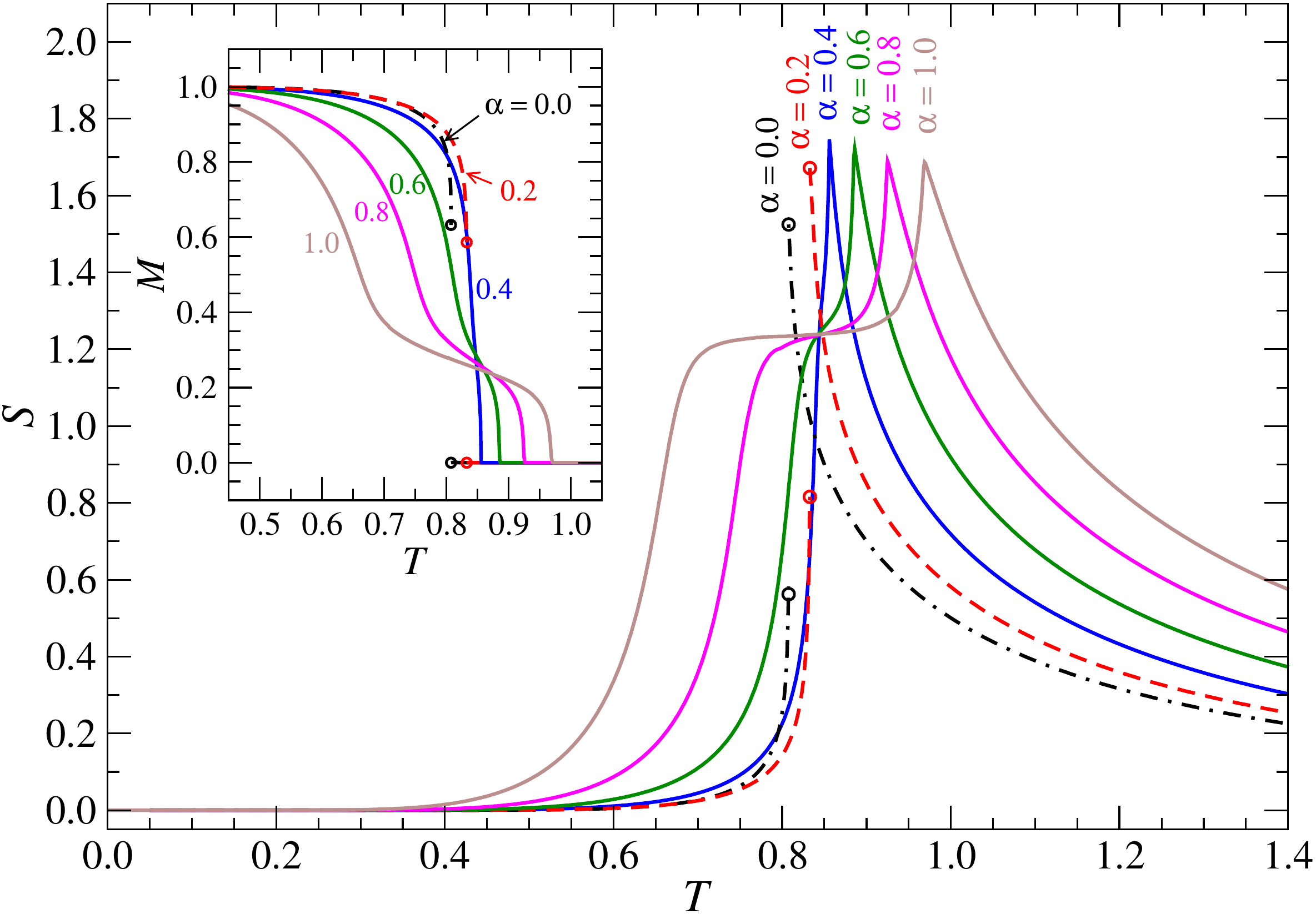}
\caption{Temperature dependence of the entanglement entropy for $\alpha=0.0$, $0.2$, $0.4$, $0.6$, $0.8$, $1.0$ and $q=6$. The inset shows magnetization $M$ versus temperature $T$ with the identical set of $\alpha$.}
\label{Fig3}
\end{figure}

Figure~\ref{Fig3} shows the entanglement entropy $S$ and the magnetization $M$ (in the inset) with respect to temperature $T$ for selected $\alpha$. The first-order transition is accompanied by a discontinuous jump in $S$ and $M$ at temperature $T_0(\alpha=0) = 1/\ln(1 + \sqrt{6}) $~\cite{FYWu} (dot-dashed lines) and $T_0(\alpha=0.2) \approx 0.833$ (dashed line). When $\alpha > 0.2$ (full lines), the typical shapes of $S$ and $M$ are signatures of the BKT phase transitions~\cite{BKT1}. We remark here that the entanglement entropy and magnetization for $\alpha = 0.4$ seem to exhibit such behavior which is typical for the continuous (second-order) phase transition. However, the detailed specification of $S$ and $M$ at $\alpha = 0.4$ results in two BKT phase transitions, as we discuss in the following.

\begin{figure}[b!]
\includegraphics[width=0.48\textwidth]{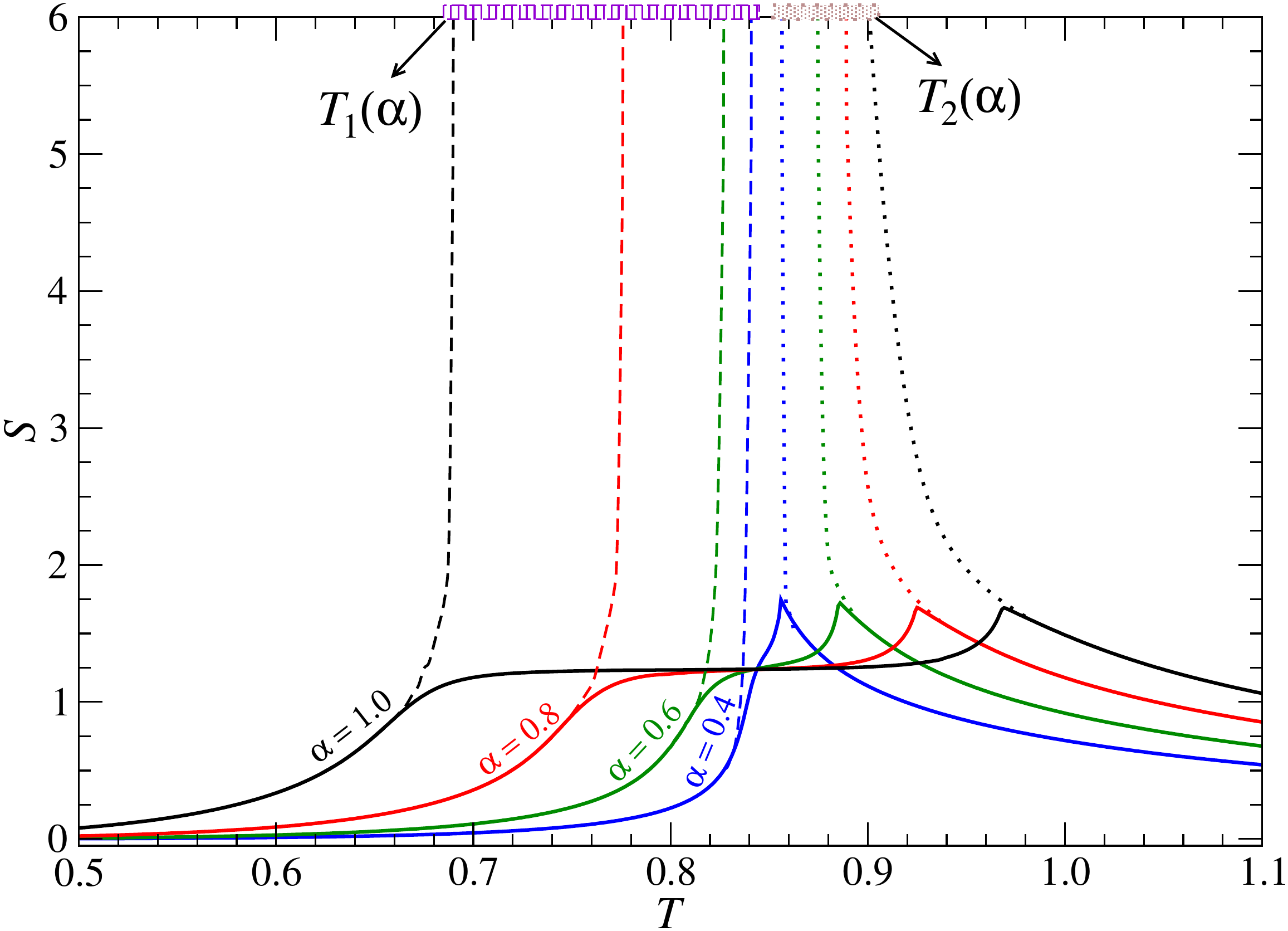}
\caption{Temperature dependence of the entanglement entropy for $\alpha = 0.4, 0.6, 0.8$, and $1.0$. In the thermodynamic limit $N \to \infty$, the extrapolation of the entanglement entropy results in its logarithmic divergence ($S_N \propto N$) in the temperature interval $T_1(\alpha) \leq T \leq T_2(\alpha)$.}
\label{Fig4}
\end{figure}

In Fig.~\ref{Fig4}, the entanglement entropy is shown in thermodynamic limit (at finite precision $m=100$) to examine the BKT phase transitions in detail. The entanglement entropy $S$ logarithmically diverges in the BKT phase~\cite{BKT1}. We have thus sketched the divergence of $S$ by the two hatched areas on top of Fig.~\ref{Fig4}. Each area consists of the BKT phase-transition temperatures  $T_1(\alpha)$ and $T_2(\alpha)$, where $S$ is infinite. The entire BKT phase is critical because the correlation length is also infinite in the entire interval $T_1(\alpha) \leq T \leq T_2(\alpha)$. The dashed and dotted lines, respectively, estimate the extrapolated critical temperatures $T_1(\alpha)$ and $T_2(\alpha)$ after taking both the limits $N\to\infty$ and $m\to\infty$.

In order to locate the tricritical point, we return back to Fig.~\ref{Fig3} which indicates the existence of $\alpha_{\rm c}$ within the interval $0.2<\alpha<0.4$. The first-order phase transition at $\alpha=0.2$ dramatically changes into a subtly curved entanglement entropy describing two BKT transitions at $\alpha=0.4$ (also visible in Fig.~\ref{Fig4}). The presence of the first-order transition at $\alpha = 0.2$ has been verified by the discontinuities of the entanglement entropy $S$ and the magnetization $M$, as depicted in Fig.~\ref{Fig3}. We calculated the latent heat $\lambda>0$ after evaluating the discontinuity of the internal energy, cf Eq.~\eqref{LH}. For this purpose, we calculated the free-energy minimum for given $\alpha$ when imposing two different boundary conditions for the system expanded into the thermodynamic~limit~\cite{PhysA, APS}. We then discuss the decreasing latent heat $\lambda \to 0$, as $\alpha$ approaches the tricritical point $\alpha_{\rm c}$ where the latent heat closes. Right above $\alpha_{\rm c}$, the two BKT transitions begin to move away from each other for $\alpha > \alpha_{\rm c}$.

Specifically, we expect that the tricritical point $T_{\rm c}(\alpha_{\rm c})$ has to be within the interval $0.2 < \alpha_{\rm c} < 0.4$ such that it satisfies the condition
\begin{equation}
\lim\limits_{\alpha\to\alpha_{\rm c}^{-}} T_0(\alpha) =
\hspace{-0.2cm}\lim\limits_{\alpha\to\alpha_{\rm c}^{+}} T_1(\alpha) =
\hspace{-0.2cm}\lim\limits_{\alpha\to\alpha_{\rm c}^{+}} T_2(\alpha) \equiv
T_{\rm c}(\alpha_{\rm c}).
\end{equation}

\begin{figure}[tb]
\includegraphics[width=0.48\textwidth]{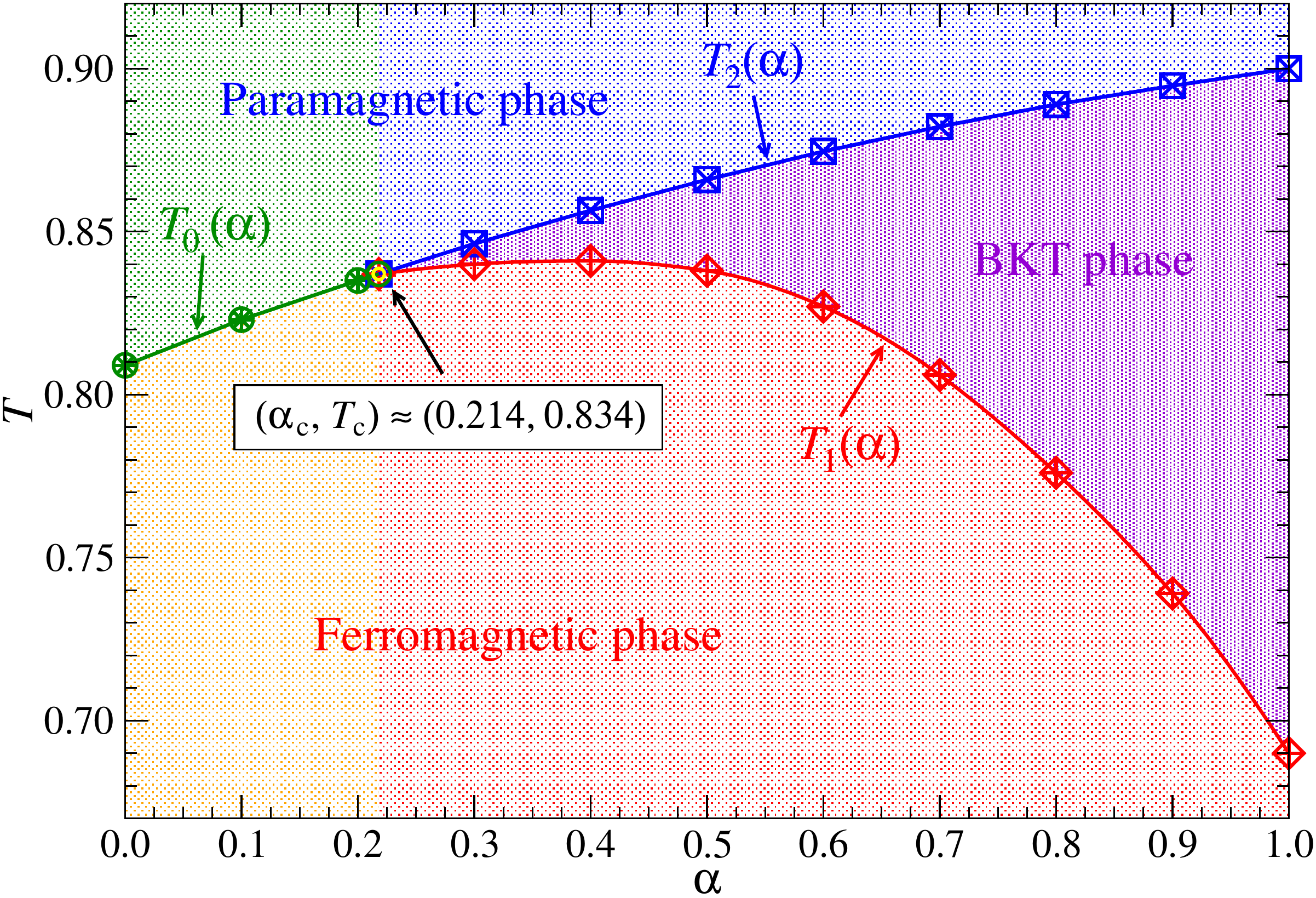}
\caption{Phase diagram of the $6$-state PC model shows the three main phases meeting at the  tricritical point $\alpha_{\rm c} \approx 0.214$ and $T_0 \approx 0.834$. The error bars (not shown) are much smaller than the symbol sizes. There are three
phase-transition lines parameterized by $\alpha$, namely, the first-order (discontinuous) $T_0 (\alpha)$, the low-temperature infinite-order (BKT) $T_1(\alpha)$, and the high-temperature infinite-order (BKT) $T_2(\alpha)$.}
\label{Fig5}
\end{figure}

\subsection{Tricritical point}

Figure~\ref{Fig5} shows the phase diagram of the PC model. The error bars are not displayed since they are much smaller than the symbol sizes. We analyze the tricritical point $\alpha_{\rm c}$ by dividing the interval $\alpha$ into the {\it left region} with $0 \leq \alpha < \alpha_{\rm c}$, where the first-order transition is present along $T_0(0 \leq \alpha < \alpha_{\rm c})$ and the {\it right region} with $\alpha_{\rm c} < \alpha \leq 1$ where two BKT transition temperatures occur, both of the infinite order along $T_1(\alpha_{\rm c} < \alpha \leq 1)$ and $T_2(\alpha_{\rm c} < \alpha \leq 1)$. The left region requires calculations of the free energy ${\cal F}$, the internal energy $U$, and the latent heat $\lambda$ to fully describe the first-order transition. The right region can be, to some extent, described by the entanglement entropy $S$ and the specific heat capacity $c$. Their shapes witness the presence of the BKT phase transitions.

{\it Left region:\ }
The position of the tricritical point $\alpha_{\rm c}$ is most reliably determined via detailed detection of $T_0(\alpha)$. To achieve the desired accuracy of $T_0(\alpha)$, we have to impose two different boundary conditions affecting the free energy ${\cal F}(\alpha)$ even in the thermodynamic limit (due to the coexistence of ordered-disordered phases in the first-order phase transitions, the boundary conditions slightly affect the free energy, including the other thermodynamic quantities, in a small region around $T_0$ also in the thermodynamic limit). The free energy is a smooth monotonically decreasing function everywhere, except at the phase transition temperature $T_0(\alpha)$. At the first-order phase-transition temperature, ${\cal F}(\alpha)$ is non-analytic such that the internal energy $U(\alpha)={\cal F}(\alpha)-T \partial{\cal F} (\alpha)/ \partial T$ becomes discontinuous at $T_0(\alpha)$. We obtain the correct $T_0(\alpha)$ by minimizing the free energy (not shown)~\cite{APS}. Consequently, we determine the remaining quantities, such as $U$, $M$, and $S$ at the correct phase transition temperature $T_0(\alpha)$.

\begin{figure}[tb]
\includegraphics[width=0.48\textwidth]{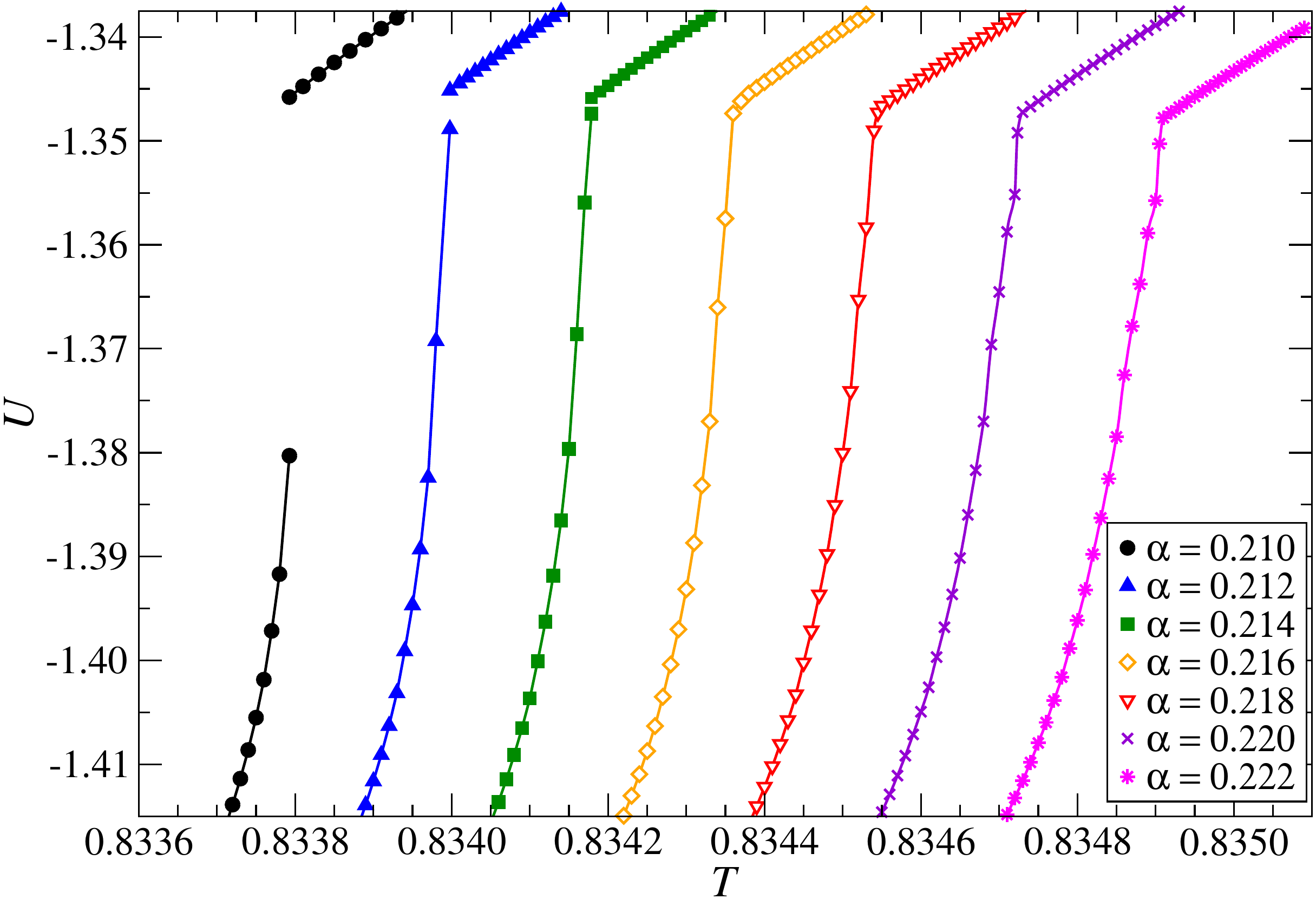}
\caption{Temperature dependence of the internal energy $U$ exhibits discontinuous jumps at $\alpha = 0.21, 0.212$, and $0.214$. The latent heat is nonzero ($\lambda>0$) if $\alpha < \alpha_{\rm c} \approx 0.214$, marked by the filled symbols.}
\label{Fig6}
\end{figure}

Once $T_0(\alpha)$ has been specified out of ${\cal F} (\alpha)$, we focus on a narrow region around the tricritical point. The internal energy $U$ (cf. Eq.~\eqref{IntEng}) shown in Fig.~\ref{Fig6} is discontinuous if $0.210 \leq \alpha \leq 0.214$ and continuous if $0.216 \leq \alpha \leq 0.222$. The discontinuity exhibits phase transition of the first order since the latent heat $\lambda>0$ when obtained from Eq.~\eqref{LH}. If approaching the limit $\alpha \to \alpha_{\rm c}^{-}$ (from left), we found out gradually closing discontinuous jump in $U$, i.e., $\lambda (\alpha) \to \lambda (\alpha_{\rm c}) = 0$, as shown in Fig.~\ref{Fig6}, noticing that $\lambda>0$ for $0\leq \alpha < \alpha_{\rm c}$. The gap closes (i.e., becomes of the second order, cf Tab.~I) at the tricritical point at $T_0\approx 0.834$ and $\alpha_{\rm c}\approx 0.214$, as marked in Fig.~\ref{Fig5}.

\begin{figure}[tb]
\includegraphics[width=0.48\textwidth]{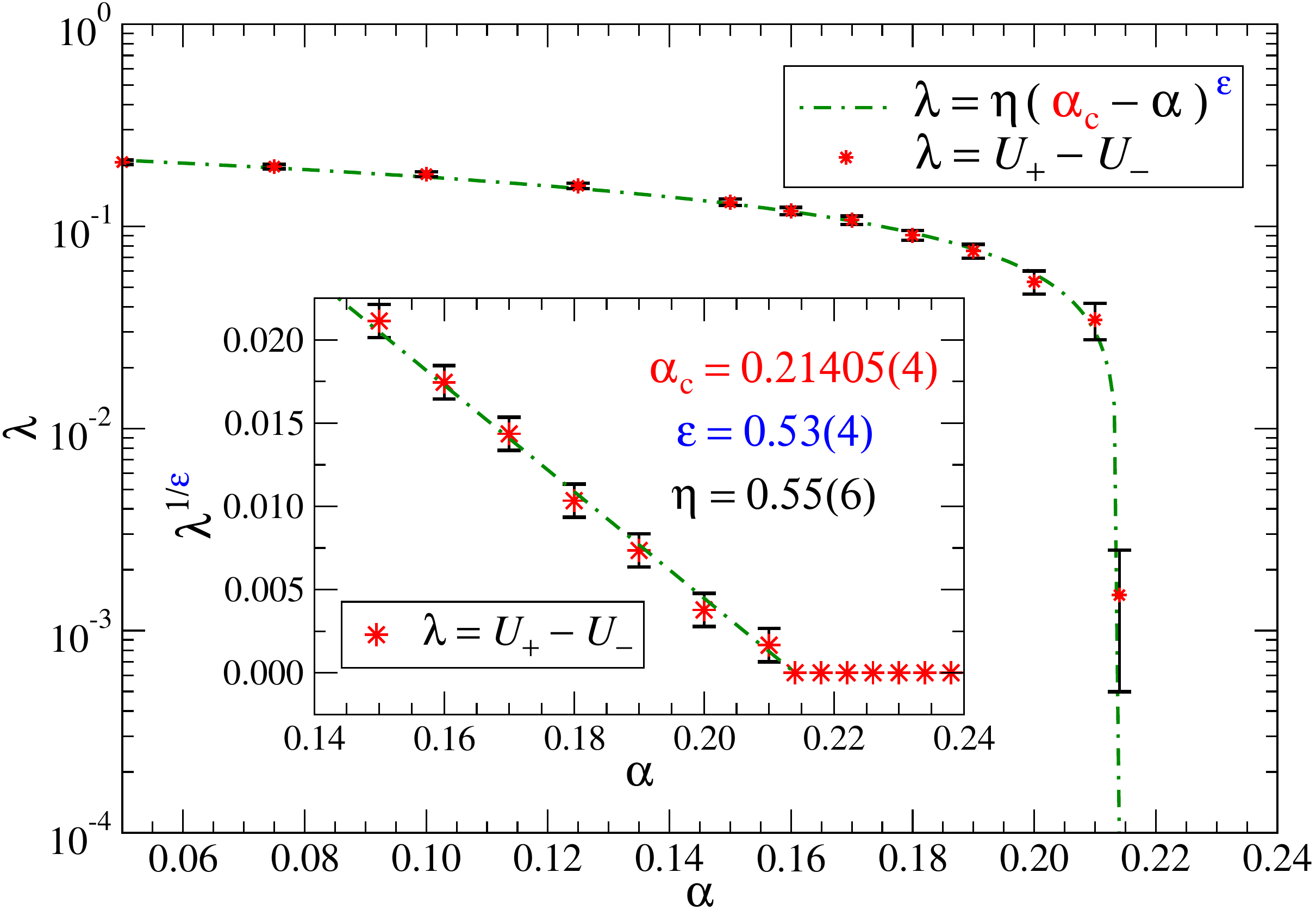}
\caption{Latent heat $\lambda$ versus $\alpha$ in the log-linear scale. The dot-dashed line is the power-law fit $\lambda(\alpha) = \eta(\alpha_{\rm c} - \alpha) ^\varepsilon$, where we determined the tricritical point at $\alpha_{\rm c}=0.21405(4)$. The inset shows a linear decay of the data after rescaling to $\lambda^{1/ \varepsilon}$.}
\label{Fig7}
\end{figure}

In the following, we estimate $\alpha_{\rm c}$ from the viewpoint of the latent-heat convergence $\lambda \to 0$ when $\alpha \to \alpha_{\rm c}$ after the careful detection of $T_0(\alpha)$ by the examination of free-energy crossover with respect to boundary conditions~\cite{PhysA}. Figure~\ref{Fig7} shows the latent heat $\lambda$ as a function of $\alpha$ in the log-linear scale. It is obvious from the graph that the latent heat drops to zero fast. Having applied the nonlinear least-square method to $\lambda(\alpha) = \eta(\alpha_{\rm c} - \alpha) ^\varepsilon$, we found the fitting parameters $\eta = 0.55(6)$, $\alpha_{\rm c} = 0.21405(4)$, and $\varepsilon = 0.53(4)$.

The inset of Fig.~\ref{Fig7} shows the $\alpha$-dependence of $\lambda^{1/\varepsilon}$ to emphasize that the latent heat with the exponent $1/\varepsilon$ linearly decreases to zero when approaching the tricritical point. Thus, we found a square-root dependence ($\varepsilon\approx0.5$), i.e., $\lambda(\alpha) \propto \sqrt{\alpha_{\rm c} - \alpha}$ for $\alpha \leq \alpha_{\rm c}$. The first-order phase transition in the PC model terminates in $\alpha_{\rm c} = 0.21405(4)$, where we expect that the second-order phase transition smoothly changes into the two BKT phase transitions, as $\alpha$ further increases above $\alpha_{\rm c}$. The error bars reflect inaccuracies in estimating $U$ and $\lambda$ with respect to the finite number of states kept $m$ after they were extrapolated to infinity (where $U$ and $\lambda$ become exact).

{\it Right region:\ }
In the BKT regime when $\alpha > \alpha_{\rm c}$, the latent heat $ \lambda (\alpha) = 0$. Notice tiny ripples in $U$ in Fig.~\ref{Fig6}, which are more visible right below the phase-transition temperature for $\alpha \gtrsim 0.218$. They indicate two inflection points leading to maxima of the specific heart $T_1$ and $T_2$  indicate the proximity of the lower-temperature BKT phase transitions at $T_1(\alpha)$ and $T_2(\alpha)$. Since the ripples are sensitive to temperature derivative, they affect the specific heat $c=\partial U/ \partial T$ which results in two peaks corresponding to the BKT transition temperatures $T_1$ and $T_2$.

\begin{figure}[tb]
\includegraphics[width=0.48\textwidth]{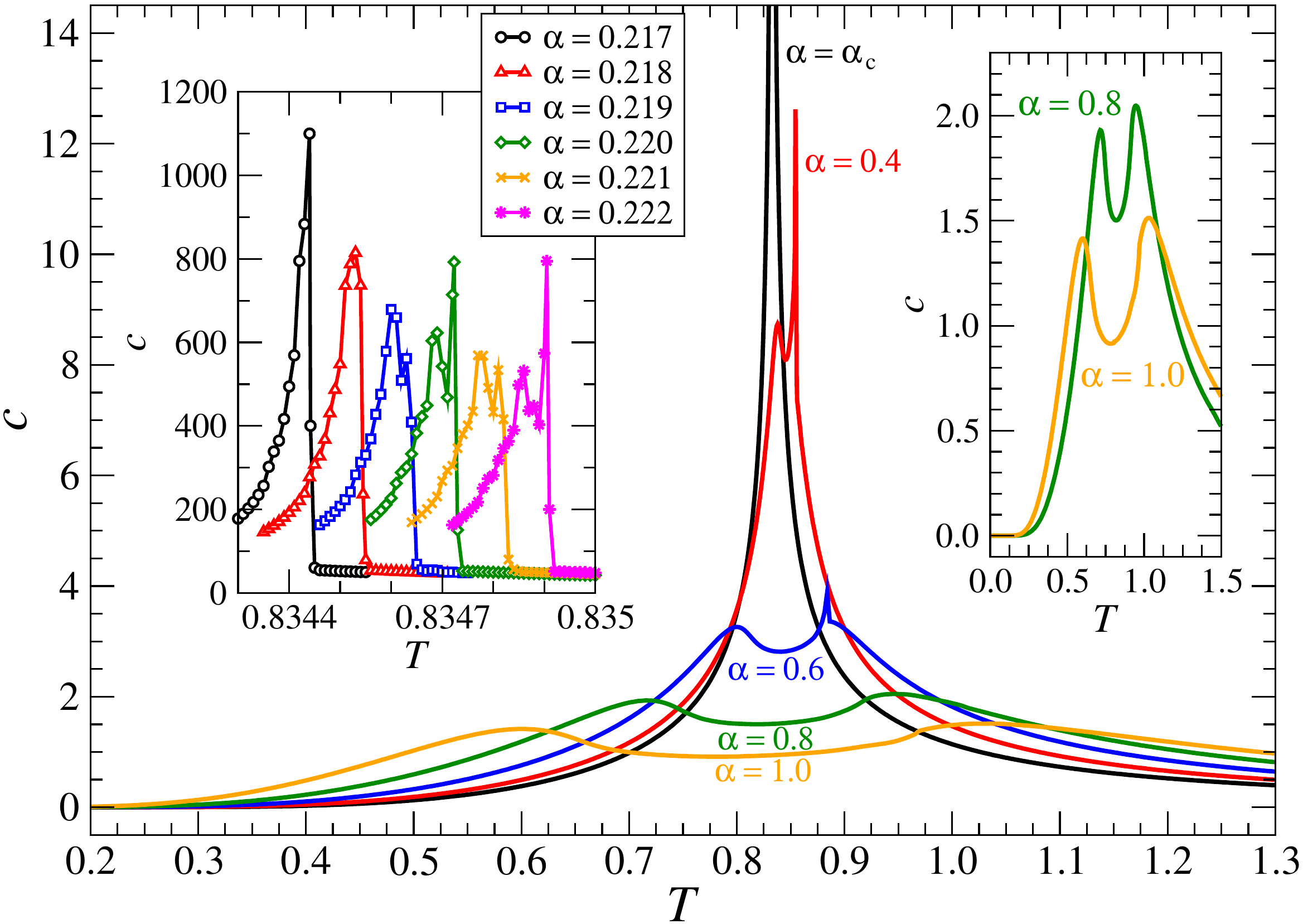}
\caption{Detailed view on the specific heat $c=\frac{\partial U}{\partial T}$ versus $T$ at and above $\alpha_{\rm c} \approx 0.214$. As $\alpha$ increases, the sharp maxima of specific heat decrease and broaden toward the known BKT transitions at $T_1(\alpha=1)$ and $T_2(\alpha=1)$. The right inset shows scaled $c$ for $\alpha=0.8$ and $1.0$. The left inset shows $c$ right above $\alpha_{\rm c}$ where the two maxima referring to $T_1(\alpha)$ and $T_2(\alpha)$ start forming.}
\label{Fig8}
\end{figure}

In Fig.~\ref{Fig8} we plotted the specific heat $c$ at $\alpha = \alpha_{\rm c}, 0.4$, $0.6$, $0.8$, and $1.0$. The sharp divergent peak at $\alpha_{\rm c}$ confirms the second order-phase transition at the tricritical point as discussed in the {\it left region} analysis. If $\alpha$ increases, the single peak at $\alpha_{\rm c}$ broadens and splits into two non-diverging peaks that separate. Their maxima decrease as $\alpha$ increases above $\alpha_{\rm c}$. The inset on the left shows the specific heat right above $\alpha_{\rm c}$ where the two maxima with high amplitude are formed, determining $T_1(\alpha)$ and $T_2(\alpha)$ (also found by the entanglement entropy). They start departing from one another, as $\alpha$ further grows. It is natural to assume that $T_2(\alpha_{\rm c}) = T_1(\alpha_{\rm c}) \equiv T_{\rm c}(\alpha_{\rm c})$ when approaching the tricritical point $\alpha \to \alpha_{\rm c}^{+}$ (from right).

As we have mentioned earlier, the $6$-state clock model ($\alpha=1$) exhibits the two BKT transitions of the infinite order. Whether the phase-transition order smoothly changes from the second order at $\alpha_{\rm c}$ to the infinite order at $\alpha=1$ or there is a direct change from the second order to the infinite from $\alpha > \alpha_{\rm c}$, we leave for future investigations, as we cannot give an unambiguous answer.

\subsection{Critical exponents at \boldmath{$T_{\rm c}(\alpha_{\rm c})$}}

To classify the universality in the tricritical point, we calculate two independent critical exponents $\beta$ and $\delta$. The exponent $\beta$ is associated with the decay of the spontaneous magnetization $M$ close to the tricritical point
\begin{equation}
	M\propto [T_{\rm c}(\alpha_{\rm c})-T ]^{\beta}\, .
\end{equation}
The second critical exponent $\delta$ is related to the dependence of the induced magnetization $M$ when imposing a constant magnetic field $h$ at $T_{\rm c}$ and $\alpha_{\rm c}$ (recalling that the critical magnetic field $h_{\rm c}=0$)
\begin{equation}
	M\propto(h - h_{\rm c})^{1/\delta}\, .
\end{equation}

For a given $m$, we first calculate magnetization unless fully converged, i.e., $M_{k,m} \to M_{\infty,m} \equiv M_m$. Subsequently, we increase $m$ until numerically feasible. Finally, we extrapolate $M_m \to M_{\infty}$. With both limits, we find the critical exponent $\beta$
\begin{equation}
\lim\limits_{k,m\to\infty} M_{k,m} = \lim\limits_{m \to \infty} \gamma_m (t_m-T)^{\beta_m}\, ,
\end{equation}
where $M_m$ is the magnetization obtained by CTMRG, and $\gamma_m$, $t_m$, and $\beta_m$ are unknown parameters obtained by the nonlinear least-square fitting. Taking the limit $m \to \infty$, we identify $t_{\infty} \equiv T_{\rm c}$ and $\beta_{\infty} \equiv \beta$.

\begin{figure}[tb]
\includegraphics[width=0.48\textwidth]{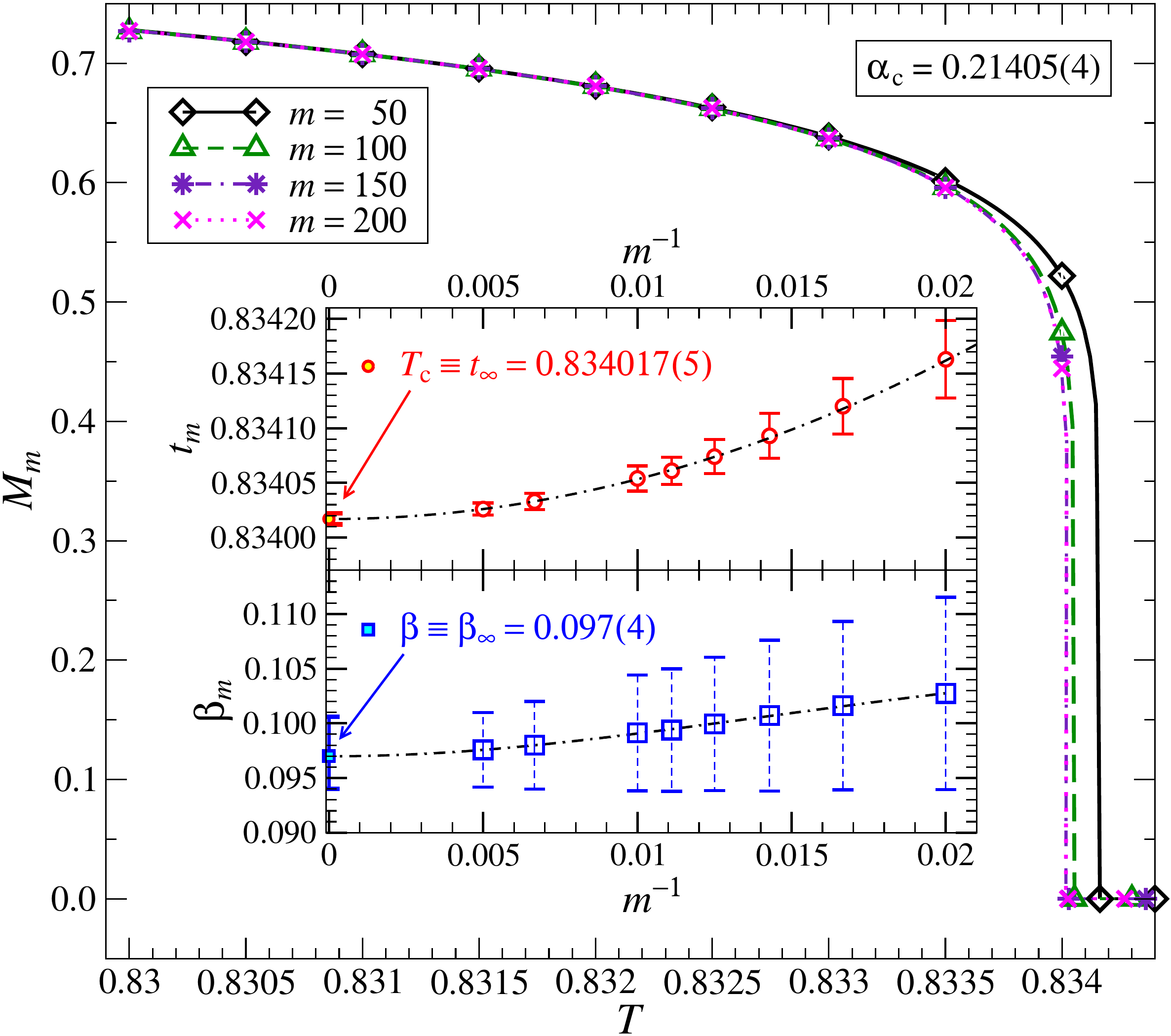}
\caption{Spontaneous magnetization $M_m$ versus temperature $T$ at the tricritical $\alpha_{\rm c}=0.21405(4)$ for $m=50, 100, 150$, and $200$ when the external magnetic field $h=0$. Inset shows two graphs with nonlinear fits of the critical temperatures $t_m$ (upper) and exponents $\beta_m$ (lower). Taking the extrapolation $m^{-1} \to 0$, we obtained $T_{\rm c} = 0.834017(5)$ and $\beta = 0.097(4)$.}
\label{Fig9}
\end{figure}

Figure~\ref{Fig9} shows the magnetization data $M_m$ for $m=50$, $100$, $150$, and $200$ at the tricritical point $\alpha_{\rm c}$ and zero magnetic field $h=0$. The inset shows the scaling of the critical temperatures $t_m$ and the critical exponents $\beta_m$ versus $1/m$ for $m=50,60,70,80,90,100,150$, and $200$. We apply the least-square method to the nonlinear expression $t_m = T_{\rm c} + \tau_m / (m + \mu_m)$ to find the critical temperature $T_{\rm c}$ (the full circle). Analogously, we estimate the critical exponent $\beta$ via fitting $\beta_m = \beta + b_m / (m + \nu_m)$ (squares). The fitting parameters $T_{\rm c}$, $\tau_m$, $\mu_m$, $\beta$, $b_m$, and $\nu_m$ are determined. In the limit $m^{-1} \to 0$, we extrapolate the critical temperature $T_{\rm c} = 0.834017(5)$ and the critical exponent $\beta = 0.097(4)$, as depicted in the insets. The error bars reflect the inaccuracy within the intervals $0.834012 \leq T \leq 0.834022$ and $0.21401 \leq \alpha \leq 0.21409$, noticing the improving error bars, as $m$ increases.

\begin{figure}[tb]
\includegraphics[width=0.48\textwidth]{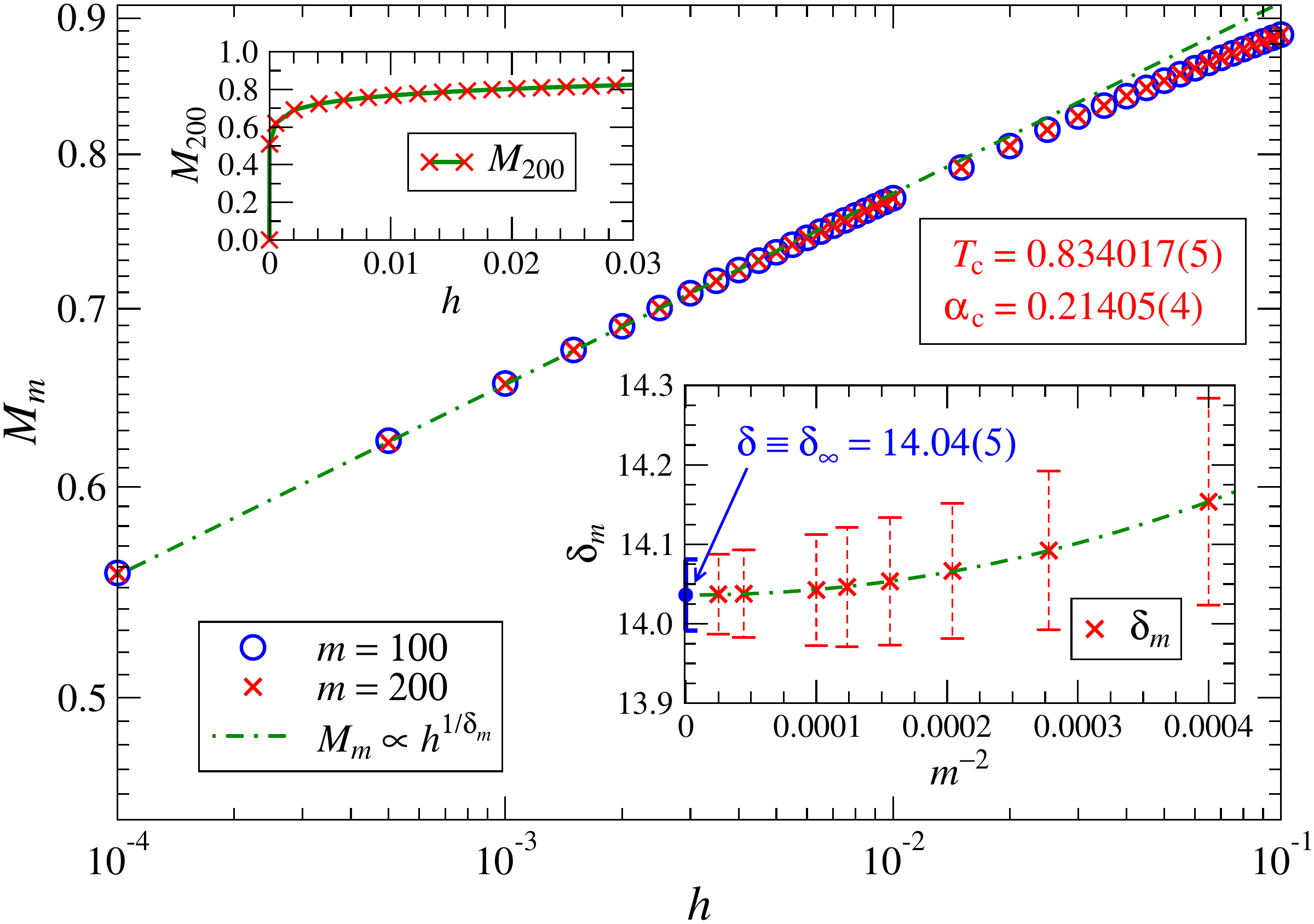}
\caption{Induced magnetization $M_m$ versus magnetic field $h$ in the log-log plot calculated at the tricritical point $T_{\rm c}=0.834017(5)$ and $\alpha_{\rm c} = 0.21405(4)$ for $m=100$ (red crosses) and $m=200$ (blue circles). The upper inset shows $M_{200}$ in the linear scale. The lower inset determines the extrapolation of the critical exponent $\delta = 14.04(5)$ in the thermodynamic limit $m^{-2} \to 0$.}
\label{Fig10}
\end{figure}

Figure~\ref{Fig10} shows the dependence of the uniform magnetic field $h$ on the magnetization $M_m$. Within the log-log plot, the power-law dependence $M_m(h) \propto h^{1/\delta_m}$ becomes linear, as shown by the dot-dashed line. There are negligible differences of $M_m$ between $m=100$ and $m=200$, which witness high numerical accuracy reached in the thermodynamic limit. The upper inset shows the magnetization $M_{200}$ in the linear scale, where we confirm a non-linear convergence of $M_m \to 0$ at the tricritical point when $h \to 0$. The lower inset shows the scaling of the second critical exponent $\delta$ determined by the non-linear fitting expression $\delta_m = \delta + d_m / (m^2 + \rho_m)$ if plotted with respect to $m^{-2}$. Here, $\delta_{\infty}$, $d_m$, and $\rho_m$ are the unknown parameters found by the least-square fitting. The extrapolation critical exponents yields $\delta_{\infty} \equiv \delta=14.04(5)$.

\section{Conclusions}

We have proposed an anisotropic deformation of the $6$-state clock model with the aim to break its rotational symmetry. For this purpose, we chose the $6$-state Potts symmetry. Our motivation was a gradual suppression of the BKT transition in the clock model competing with the discontinuous Potts-like transition. The asymmetric deformation of the competing interactions gives rise to a tricritical point, where first and infinite orders meet.

Using the CTMRG method, we have constructed the phase diagram of the $6$-state PC model, as depicted in Fig.~\ref{Fig5}. We determined the tricritical point at $\alpha_{\rm c}=0.21405(4)$ and $T_{\rm c}=0.834017(5)$, where the second-order transition occurs. The associated critical exponents are $\beta = 0.097(4)\approx \frac{1}{10}$ and $\delta = 14.04(5)\approx14$. They fall within the limits of the exactly known exponents for the $q=2,3,4$ Potts and clock models. In particular, $\beta=\frac{1}{8}$ and $\delta=15$ characterize both the $2$-Potts and $2$-clock models, whereas $\beta=\frac{1}{9}$ and $\delta=14$ belong to the $3$-Potts and clock models. However, the $4$-Potts model exhibits $\beta=\frac{1}{12}$ and $\delta=15$ while for the $4$-clock model, we get $\beta=\frac{1}{8}$ and $\delta=15$~\cite{FYWu, Chen}.

The phase diagrams of any $q$-state PC models with finite $q>6$ are expected to be qualitatively similar to the current study ($q=6$), due to the similarities of the models at $\alpha = 0$ and $\alpha = 1$. This is evident from the phase diagram in Fig.~\ref{Fig1}. However, if $q=5$, we observe a weak first-order transition in the close vicinity of narrow BKT transitions, which may lead to lowered numerical accuracy. Moreover, the phase-diagram structure for $q=5$ differs significantly since $T_0(0) < T_1(1) < T_2(1)$, whereas $T_1(1) < T_0(0) < T_2(1)$ for $q\geq6$. The case $q=5$ deserves a detailed investigation.

The question of whether the infinite-order phase transition persists for $\alpha_{\rm c} < \alpha \leq 1$ or continuously varies from the second-order at $\alpha = \alpha_{\rm c}$ towards the infinite order at $\alpha = 1$ (via third, fourth, etc. order transition) is left unanswered. Having introduced the anisotropic deformation of the spin rotational symmetry in the $q \geq 6$-state clock models by the Potts interaction gives rise to the $q$-state PC model. The PC model weakens the BKT-type transition by continuously decreasing from $\alpha = 1$ to $\alpha = \alpha_{\rm c}$ which makes this transition to be numerically tractable. This work also opens novel viewpoints on various magnetic irregularities when observing anisotropic spin behavior experimentally. We intend to resolve the question in a future study due to the numerical limitations of the CTMRG method.

\section*{Acknowledgments}
We would like to thank Tomotoshi Nishino for valuable discussions.
This work was partially funded by Agent\'{u}ra pre Podporu V\'{y}skumu a V\'{y}voja (No. APVV-20-0150), Vedeck\'{a} Grantov\'{a} Agent\'{u}ra M\v{S}VVa\v{S} SR a SAV (VEGA Grant No. 2/0156/22), and Joint Research Project SAS-MOST 108-2112-M-002-020-MY3.

\end{document}